\documentclass[12pt,a4paper]{article}
\usepackage[english]{babel}
\usepackage[cp1251]{inputenc}
\usepackage{latexsym,amssymb,amsmath,amsthm} 
\usepackage{graphicx}
\usepackage{hyperref}
\usepackage{amsfonts}

\def\bb{\begin{equation}}
\def\ee{\end{equation}}
\def\pt{\partial}

\def\ve{\varepsilon}

\begin{document}

\author{Nataliya Gorbatova\footnote{Ufa State Aviation Technical University},\quad Oleg Kiselev\footnote{Institute of Mathematics of USC RAS, e-mail: ok@ufanet.ru},\quad
Sergei Glebov\footnote{Ufa State Petroleum Technical University, e-mail: sg@anrb.ru}}

\title{Finite amplitude waves under a small resonant driving force}
 
\date{}
\maketitle

\begin{abstract}
We construct a special asymptotic solution for the forced Boussinesq equation. The perturbation is small and oscillates with a slowly varied frequency. The slow passage through the resonance generates waves with the finite amplitude. This phenomenon is described in details. 
\end{abstract}


\section*{Introduction}
\addcontentsline{toc}{section}{Introduction}

The Boussinesq equation is one of  fundamental equations of the mathematical physics. It naturally appears when one investigates the propagation of long waves into nonlinear media.  The Boussinesq equation is the basic model for the description of surface waves \cite{graig},\cite{kano},\cite{kano1}; cross-axis oscillations of the balks and rods. It describes wave phenomenons in plasma and optical fibers \cite{petviashvili},\cite{wisem}. 
\par
We consider the forced Boussinesq equation
$$
U_{tt}-U_{xx}+a(U_x)^2_x + b U_{xxxx} = \ve^2 {\cal F}.
$$
The external perturbation has the different meaning for various applications: periodic influences on rods or an external pumping in optical media. The characteristic feature is the periodic nature of the external perturbation. Here we consider the such type of the perturbation
\bb
U_{tt}-U_{xx}+a(U_x)^2_x +\ve\gamma U_{xxxx} = \ve^2 f(\ve x)\exp\{iS(\ve^2 x, \ve^2 t)/\ve^2\}+\hbox{c.c.}, \label{bouss}
\ee
where $\ve$ is a small positive parameter; $a, \gamma$ are constants. The amplitude $f(y)$ of the driving force is bounded and rapidly vanished function, when $\vert y \vert \to \infty$. 
We investigate the simplest case $S = S(\ve^2 t) = (\ve^2 t)^2/2$. It allows us to avoid the complicated calculations and saves the essence of the phenomenon.
\par
Before the resonance oscillations are realized under the perturbation. In a neighborhood of the curve $t=0$ the driving force becomes resonant and it leads to the change of the solution behaviour. The solution contains two waves of the finite amplitude after the slow passage through the resonance. In contrast to known papers related to the slow passage through  resonance \cite{ok-sg1}, \cite{ok-sg2}, \cite{ok-sg3} in this situation we obtain  waves with the amplitude of $O(1)$.  This large increase of the order of the solution follows from the resonance on the zero harmonic.
\par
The paper has the following structure. In the first section we briefly describe the result of the article and explain the observed phenomenon. The second section presents results of numerical simulations. Then we realize analytical asymptotic constructions. The forced oscillating solution of order $\ve^2$ is constructed in the third section. The forth section contains the constructions in a neighborhood of the resonant curve. The behaviour of the solution after the passage through the resonance are described in the fifth section. All constructed asymptotic expansions are matched in the manner of \cite{Ilin}.
\par

\section{Main result}

The formal asymptotic solution for (\ref{bouss}) is constructed in the certain domain. This domain is divided on several subdomains and covers the resonant line $t=0$. Our special solution has a different representation in  each subdomain. 
\par
The forced oscillations 
$$
U(x,t,\ve) \sim \ve^2 u_2(\ve x, \ve^2 t) \exp\{i(\ve t)^2/2\}
$$
describe the solution behaviour when $t<0$. In a neighborhood of the resonant curve $t=0$ the asymptotic solution is represented by 
$$
U(x,t,\ve) \sim  w_0(\ve x,\ve t)
$$
 The dynamics of the leading-order term $w_0$ is described in terms of the Fresnel integral.  After the passage through the resonance the solution becomes $O(1)$
$$
U(x,t,\ve) \sim v_0^+(\ve t +\ve x,\ve^2 t)+v_0^-(\ve t -\ve x,\ve^2 t). 
$$ 
This postresonant domain corresponds to $t>0$. The leading-order terms $v_0^+, v_0^-$ of the solution  are  determined from the pair of the Hopf equations (\ref{Hopf}). All these representations are matched.
\par

\section{Numerical simulations}

In this section we present results of numerical simulations. 
They completely justify our asymptotic formulas and were done under 
the following conditions: $\ve=0.05$ and the perturbation looks like 
$\displaystyle\frac{1}{1+(\ve x)^4}\cos\left(\frac{(\ve^2 t)^2}{2\ve^2}\right)$. 
The figure shows the generation of two waves of the finite amplitude of $O(1)$. 
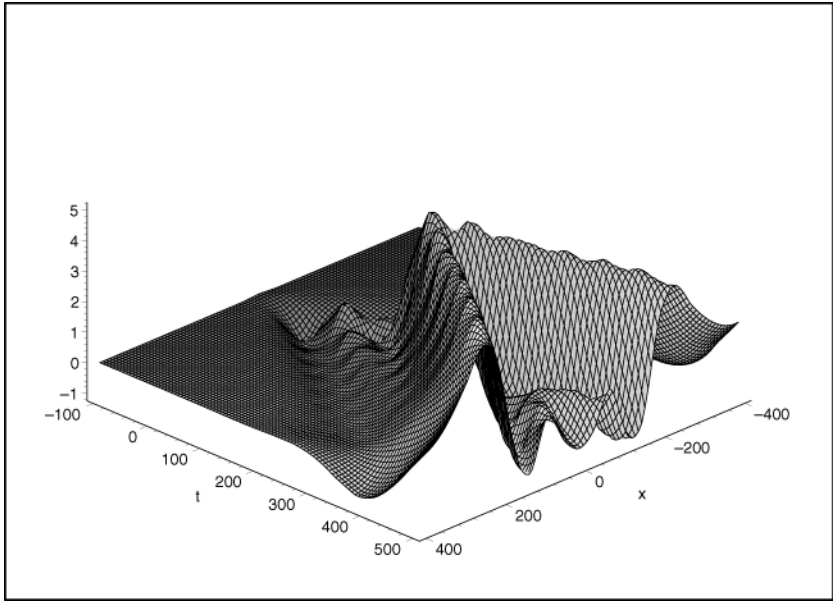
\begin{center}
\includegraphics[width=11cm]{waves.ps}
\end{center}

\section{The first external expansion}\label{theFirstExternalExpansion}

In this section we construct the asymptotic solution of (\ref{bouss}) that corresponds to forced oscillations. This solution has the order  $\ve^2$ and oscillates under the  perturbation.

\subsection{Equations for coefficients}

The WKB solution has the form
\begin{eqnarray}
U(x,t,\ve) &=& \Big[\ve^2 u_2(x_1,x_2,t_2) + \ve^4 u_4(x_1,x_2,t_2) \nonumber \\
&+& \ve^6 u_6(x_1,x_2,t_2) \Big]\exp\{iS(t_2)/\ve^2\} + c.c.,
 \label{firstExternalAnzats}
\end{eqnarray}
where $x_m = \ve^m x, t_m = \ve^m t$ for $m=1,2$.
\par
Substitute this representation into (\ref{bouss}) and gather the terms with the same order with respect to $\ve$. It yields the series of the algebraic equations
\par
\bb
(S')^2 u_2 = -f, \label{ext2}
\ee
\bb
(S')^2 u_4 =  2i S'\pt_{t_2}u_2  + iS''u_2 - \pt_{x_1}^2u_2, \label{ext4}
\ee
\bb
(S')^2 u_6 = iS''u_4 + 2iS'\pt_{t_2}u_4 +\pt_{t_2}^2u_2 - \pt_{x_1}u_4 \label{ext6}
\ee
Note that $S' = t_2$ in our model case. It allows one to obtain the exact representation for   $u_n$.
\bb
u_2 = - \frac{f}{t_2^2},\label{sext2}
\ee
\bb
u_4 = \frac{3if+\pt_{x_1}^2f}{t_2^4}, \label{sext4}
\ee
\bb
u_6 = \frac{15f -10i\pt_{x_1}^2f - \pt_{x_1}^4f}{t_2^6}. \label{sext6}
\ee

\subsection{The domain of validity for external expansion (\ref{firstExternalAnzats})}

One can see that equations (\ref{ext2}) - (\ref{ext6}) are not solvable, when $S'=0$. Moreover the representation (\ref{firstExternalAnzats}) loses the asymptotic property in a neighborhood of the curve 
\bb
l[S]=S'=t_2 = 0.\label{resonantCurve}
\ee
Using (\ref{sext2}) - (\ref{sext6}) we obtain the representation of the solution 
in a neighborhood of the resonant curve 
\begin{eqnarray}
U(x,t,\ve) &=& \left[\ve^2 \left(- \frac{f}{t_2^2}\right)  + \ve^4\left(\frac{3if+\pt_{x_1}^2f}{t_2^4}\right)\right. \nonumber \\
&+&\left.\ve^6 \left(\frac{15f -10i\pt_{x_1}^2f - \pt_{x_1}^4f}{t_2^6}\right)\right]
\exp\Big\{i\frac{t_1^2}{2}\Big\} + c.c. \label{asymptoticsAs0-}
\end{eqnarray}
Representation  (\ref{firstExternalAnzats}) saves the asymptotic property when 
$$
\ve \max_{x_1,x_2,t_2}\vert U_{n+1}\vert = o \left(\max_{x_1,x_2,t_2}\vert U_{n}\vert\right), \quad \ve \to 0.
$$ 
It yields the domain of validity 
$$
-t_2 \gg \ve.
$$

\section{The internal expansion}\label{theInternalExpansion}

In this section we construct the asymptotic expansion of the solution in a neighborhood of resonant curve (\ref{resonantCurve}). Here we use new scaled variables $x_1$ and $t_1$ and construct the solution of the form
\bb
U(x,t,\ve) =  w_0(x_1,t_1)+ \ve w_1(x_1,t_1) + c.c. \label{internalExpansion}
\ee

\subsection{Equations for coefficients}

Let us to substitute (\ref{internalExpansion}) into (\ref{bouss}) and gather the terms of the same order with respect to $\ve$. 
\par
The leading-order term satisfies to
\bb
\pt_{t_1}^2w_0 - \pt_{x_1}^2w_0 = f(x_1)\exp\{it_1^2\} + c.c. \label{insideLeadingOrderEquation}
\ee
The first-order correction term is determined from 
\bb
\pt_{t_1}^2w_1 - \pt_{x_1}^2w_1 = -2a\left[\pt_{x_1}w_0\pt_{x_1}^2w_0 + \pt_{x_1}w_0^*\pt_{x_1}^2w_0 \right] + c.c. \label{insideFirstCorrectionEquation}
\ee

\subsection{Solutions and asymptotics}

First of all we solve equation (\ref{insideLeadingOrderEquation}). The equation for characteristics is
\bb
\left(\frac{\pt \omega}{\pt t_1}\right)^2 - \left(\frac{\pt \omega}{\pt x_1}\right)^2 =0, \label{characteristics}
\ee
Introduce new characteristic variables 
$$
y = t_1 + x_1, \quad z = t_1 - x_1.
$$ 
Then we obtain the equation for $v(y,z)=w_0(x_1,t_1)$
\bb
4\pt_{yz}^2 v = f\left(\frac{y-z}{2}\right)\exp\left\{\frac{i}{2} \Big(\frac{y+z}{2}\Big)^2\right\} + c.c. \label{equationInCharacteristics}
\ee
The general solution of (\ref{equationInCharacteristics}) is represented by
\bb
4v(y,z) = V_1(y) + V_2(z) + \int_{-\infty}^z \int_{-\infty}^y f\left(\frac{\alpha-\beta}{2}\right)\exp\left\{\frac{i}{2} \Big(\frac{\alpha+\beta}{2}\Big)^2\right\}d\alpha d\beta+c.c., \label{w0solution}
\ee
where $V_1(y)$ and $V_2(z)$ are arbitrary continuously differentiable functions. 
\par
To determine these functions $V_1(y)$ and $V_2(z)$ we match this leading-order solution with expansion (\ref{firstExternalAnzats}). Let us to return to original variables
\begin{eqnarray}
w_0(t_1&+&x_1,t_1-x_1) = W^{+}_0(t_1+x_1) + W^{-}_0(t_1-x_1) \label{w0solutionTX}\\
&+&\frac{1}{4} \int_{-\infty}^{t_1-x_1} \int_{-\infty}^{t_1+x_1} f\left(\frac{\alpha-\beta}{2}\right)\exp\left\{\frac{i}{2} \Big(\frac{\alpha+\beta}{2}\Big)^2\right\}d\alpha d\beta+c.c.,\nonumber
\end{eqnarray}
and construct the asymptotics of (\ref{w0solutionTX}), when $t_1\to -\infty$. The desired asymptotics is obtained by integration by parts 
\bb
w_0 = \exp\Big\{\frac{it_1^2}{2}\Big\}\sum_{k=1}^{\infty}w_{0,2k}t_1^{-2k} + W^{+}_0(t_1+x_1) + W^{-}_0(t_1-x_1), \label{asymptoticsLOT-infty}
\ee
where $w_{0,2} = -f(x_1), w_{0,4} = 3if + \pt_{x_1}^2f$
\par
Matching asymptotics (\ref{asymptoticsLOT-infty}) and (\ref{asymptoticsAs0-}) allows one to determine terms outside the integral in (\ref{w0solutionTX}). It yields $W^{+}_0(t_1+x_1) = W^{-}_0(t_1-x_1) =0$.
\par
The solution of (\ref{insideFirstCorrectionEquation}) has the form
\bb
w_1 = \frac{1}{4}\int_{-\infty}^{t_1-x_1} \int_{-\infty}^{t_1+x_1} H(\alpha,\beta) d\alpha d\beta + W_1^{+}(t_1+x_1) + W_1^{-}(t_1-x_1), \label{solution1CT}
\ee
where
\bb
H(\alpha, \beta) = -2a\left(\pt_{x_1}w_0\pt_{x_1}^2w_0 + \pt_{x_1}w_0^*\pt_{x_1}^2w_0\right).\label{RHS1}
\ee
After substituting (\ref{asymptoticsLOT-infty}) into (\ref{RHS1}) we obtain
\begin{eqnarray}
w_1 &=& \frac{1}{4}\int_{-\infty}^{t_1-x_1} \int_{-\infty}^{t_1+x_1}\Big[\left(\frac{A(\frac{\alpha-\beta}{2})}{(\frac{\alpha+\beta}{2})^4} + O\left(\frac{\alpha+\beta}{2}\right)^{-6} \right)\exp\Big\{i\left(\frac{\alpha+\beta}{2}\right)^2\Big\} \nonumber\\
&+& \frac{A(\frac{\alpha-\beta}{2})}{(\frac{\alpha+\beta}{2})^4} + O\left(\frac{\alpha+\beta}{2}\right)^{-6} \Big]d\alpha d\beta + W_1^+ +W_1^- +c.c.
\end{eqnarray} 
Taking the integral by parts gives
\begin{eqnarray}
w_1 = \left(\frac{A(x_1)}{t_1^6}+O(t_1^{-8})\right)\exp\{it_1^2\} + \left(\frac{A(x_1)}{6t_1^2}+O(t_1^{-4}) \right)\nonumber \\
 + W_1^+(t_1+x_1)+ W_1^-(t_1-x_1) +c.c. \label{asymptotics1OT-infty}
\end{eqnarray}
In these formulas we keep the leading-order terms of expansions only. This rough approach is sufficient to obtain the domain of validity of (\ref{internalExpansion}) when $t_1 \to -\infty$ and match the solution with representation (\ref{asymptoticsAs0-}).
\par
Matching with (\ref{asymptoticsAs0-}) yields $W_1^+(t_1+x_1) = W_1^-(t_1-x_1) = 0$.
\par 
Formulas (\ref{asymptoticsLOT-infty}) and  (\ref{asymptotics1OT-infty}) allow us to determine the domain of validity for (\ref{internalExpansion}) when $t_1 \to -\infty$. The condition
$$
\ve \max_{x_2,t_2}\vert w_{n+1}\vert = o \left(\max_{x_2,t_2}\vert w_{n}\vert\right), \quad \ve \to 0
$$
is equivalent $\ve \ll 1$. It means that expansion  (\ref{internalExpansion}) is valid for $t_1 <0$.

\subsection{Asymptotics as $t_1\to+\infty$}

In this subsection we construct asymptotics of $w_n$ as $t_1\to+\infty$ and determine the domain of validity for internal expansion (\ref{internalExpansion}).
\par
Fist of all we investigate the leading-order term $w_0$. As was shown above the coefficient $w_0$ is represented by formula (\ref{w0solution}). This formula can be written as follows 
\begin{eqnarray}
w_0 &=& \frac{1}{4}\int_{-\infty}^{\infty}\int_{-\infty}^{\infty}f\left(\frac{\alpha - \beta}{2}\right)\exp\left\{i\left(\frac{\alpha + \beta}{2}\right)^2\right\}d\alpha d\beta  \nonumber\\ &-&\frac{1}{4}\int_{-\infty}^{\infty}\int_{t_1 + x_1}^{\infty}f\left(\frac{\alpha - \beta}{2}\right)\exp\left\{i\left(\frac{\alpha + \beta}{2}\right)^2\right\}d\alpha d\beta \nonumber\\
&-&\frac{1}{4}\int_{t_1 - x_1}^{\infty}\int_{-\infty}^{\infty}f\left(\frac{\alpha - \beta}{2}\right)\exp\left\{i\left(\frac{\alpha + \beta}{2}\right)^2\right\}d\alpha d\beta \nonumber \\
&+& \frac{1}{4}\int_{t_1 - x_1}^{\infty}\int_{t_1 + x_1}^{\infty}f\left(\frac{\alpha - \beta}{2}\right)\exp\left\{i\left(\frac{\alpha + \beta}{2}\right)^2\right\}d\alpha d\beta + c.c. \nonumber
\end{eqnarray}
After evident denotations we obtain
\bb
w_0 =C + w_0^+(t_1+x_1) + w_0^-(t_1-x_1) + \frac{R(x_1,t_1)}{t_1^2}\exp\left\{i\frac{t_1^2}{2}\right\}+c.c. \label{representationLOT+infty}
\ee
The last term of (\ref{representationLOT+infty})  decreases when $\vert t_1\vert \to \infty$. This representation is obtained after single integration by parts.
\par
To obtain the asymptotics of $w_1$ when $t_1 \to +\infty$ let us to analyze right-hand side (\ref{RHS1}) . It contains  several character terms
\begin{eqnarray}
f_1 &=& g^+(t_1+x_1) + g^-(t_1-x_1) + g(t_1+x_1,t_1-x_1)+ \frac{F_1(x_1,t_1)}{t_1^{2}}\exp\{it_1^2/2\} \nonumber \\
 &+& \frac{F_2(x_1,t_1)}{t_1^{2}}\exp\{-it_1^2/2\} + \frac{F_3(x_1,t_1)}{t_1^{4}}\exp\{it_1^2\} + \frac{F_4(x_1,t_1)}{t_1^{4}},\qquad \label{asyptoticsRHS1CT}
\end{eqnarray}
where
$$
g^\pm(t_1\pm x_1)  = -2a\Big(\pt_{x_1}w_0^\pm \pt_{x_1}^2 w_0^\pm + \pt_{x_1}w_0^{*\pm} \pt_{x_1}^2 w_0^\pm\Big),
$$
\begin{eqnarray*}
g(t_1+ x_1,t_1-x_1)  &=& -2a\Big(\pt_{x_1}w_0^+\pt_{x_1}^2 w_0^- + \pt_{x_1}w_0^-\pt_{x_1}^2 w_0^+ \\
&+& \pt_{x_1}w_0^{*+} \pt_{x_1}^2 w_0^- + \pt_{x_1}w_0^{*-} \pt_{x_1}^2 w_0^+\Big),
\end{eqnarray*}
\begin{eqnarray*}
F_1(x_1,t_1)&=&-2a\Big(\pt_{x_1}R\Big[\pt_{x_1}^2w_0^++\pt_{x_1}^2w_0^-\Big] \\ 
&+& \pt_{x_1}^2R\Big[\pt_{x_1}w_0^+\pt_{x_1}w_0^-+\pt_{x_1}w_0^{*+}\pt_{x_1}w_0^{*-}\Big]\Big),
\end{eqnarray*}
$$
F_2(x_1,t_1)=-2a\pt_{x_1}R^*\Big[\pt_{x_1}^2w_0^++\pt_{x_1}^2w_0^-\Big],
$$
$$
F_3(x_1,t_1)=-2a\pt_{x_1}R\pt_{x_1}^2R,
$$
$$
F_4(x_1,t_1)=-2a\pt_{x_1}R^*\pt_{x_1}^2R.
$$
After integration we obtain the following asymptotics
\begin{eqnarray}
w_1 = (t_1-x_1)w_1^+ + (t_1+x_1)w_1^- + w_1^\pm + \frac{\widetilde F_1}{t_1^4}\exp\{it_1^2/2\}  \nonumber \\
 + \frac{\widetilde F_2}{t_1^4}\exp\{-it_1^2/2\} + \frac{\widetilde F_3}{t_1^6}\exp\{it_1^2\} + \frac{\widetilde F_4}{t_1^2} + c.c.
\end{eqnarray}
Here we denote
$$
w_1^+ = \frac{1}{4}\int_{-\infty}^{t_1+x_1}g^+(y)dy,
$$
$$
w_1^- = \frac{1}{4}\int_{-\infty}^{t_1-x_1}g^-(y)dy,
$$
$$
w_1^\pm = \frac{1}{4}\int_{-\infty}^{t_1-x_1}\int_{-\infty}^{t_1+x_1}g(y,z)dydz,
$$
\begin{eqnarray*}
\frac{\widetilde F_1(x_1,t_1)}{t_1^4}\exp\{it_1^2/2\}  &=& \frac{1}{4}\int_{-\infty}^{t_1-x_1}\int_{-\infty}^{t_1+x_1}F_1\left(\frac{y+z}{2},\frac{y-z}{2}\right)\\
&\times&
\exp\left\{\frac{i}{2}\left(\frac{y+z}{2}\right)^2\right\}\left(\frac{y+z}{2}\right)^{-2}dy dz,
\end{eqnarray*}
\begin{eqnarray*}
\frac{\widetilde F_2(x_1,t_1)}{t_1^4}\exp\{-it_1^2/2\}  &=& \frac{1}{4}\int_{-\infty}^{t_1-x_1}\int_{-\infty}^{t_1+x_1}F_2\left(\frac{y+z}{2},\frac{y-z}{2}\right)\\
&\times&
\exp\left\{-\frac{i}{2}\left(\frac{y+z}{2}\right)^2\right\}\left(\frac{y+z}{2}\right)^{-2}dy dz,
\end{eqnarray*}
\begin{eqnarray*}
\frac{\widetilde F_3(x_1,t_1)}{t_1^6}\exp\{it_1^2\}  &=& \frac{1}{4}\int_{-\infty}^{t_1-x_1}\int_{-\infty}^{t_1+x_1}F_3\left(\frac{y+z}{2},\frac{y-z}{2}\right)\\
&\times&
\exp\left\{i\left(\frac{y+z}{2}\right)^2\right\}\left(\frac{y+z}{2}\right)^{-4}dy dz,
\end{eqnarray*}
\begin{eqnarray*}
\frac{\widetilde F_4(x_1,t_1)}{t_1^2} = \frac{1}{4}\int_{-\infty}^{t_1-x_1}\int_{-\infty}^{t_1+x_1}F_4\left(\frac{y+z}{2},\frac{y-z}{2}\right)
\left(\frac{y+z}{2}\right)^{-4}dy dz.
\end{eqnarray*}
To determine the domain of validity for (\ref{internalExpansion}) we use the inequality
$$
\ve \max_{x_1,t_1}\vert w_{n+1}\vert = o \left(\max_{x_1,t_1}\vert w_{n}\vert\right), \quad \ve \to 0.
$$
It yields 
$$
\vert t_1 \pm x_1 \vert \ll \ve^{-1}.
$$

\section{The second external expansion}

In this section we construct the  second external expansion that describes the behaviour of the solution after the slow passage through the resonance. 

\subsection{Equations deriving}

We construct the solution of the form
\begin{eqnarray}
U(x,t,\ve) &=& C+ v_0^+(t_1+x_1,t_2)+v_0^-(t_1-x_1,t_2)    \nonumber \\
&+& \ve \Big[v_1^+(t_1+x_1,t_2) + v_1^-(t_1-x_1,t_2)  + v_1(t_1+x_1,t_1-x_1,t_2) \Big] \nonumber \\
&+& \ve^2\Big[ v_{2,S}(x_1,t_2)\exp\{iS/\ve^2\} + v_2^+(t_1+x_1,t_2)  \nonumber \\
&+&v_2^-(t_1-x_1,t_2)+ v_2(t_1+x_1,t_1-x_1,t_2) \Big]+c.c.
\label{secondExternalExpansion}
\end{eqnarray}
The coefficients of (\ref{secondExternalExpansion}) are determined recurrently. 
The function $v_{2,S}$ is determined from the traditional algebraic equation
\bb
-(S')^2 v_{2,S} = f. \label{2extAlgebraic}
\ee
On the next step we obtain the differential equation
\bb
2\pt_{t_2\zeta}v_0^+ + 2a \pt_{\zeta}v_0^+\pt_{\zeta}^2v_0^+ + 2\pt_{t_2\eta}v_0^- + 2a \pt_{\eta}v_0^-\pt_{\eta}^2v_0^- + 4\pt_{\zeta\eta}^2v_1 = 0, \label{beforeHopf}
\ee
where $\zeta = t_1 + x_1$ and $\eta = t_1 - x_1$.
Consider this equation as equation for $v_1$. It easy to see there is a special structure of the right-hand side of the equation
$$
\pt_{\zeta\eta}^2v_1 = f^+(\zeta,t_2) + f^+(\eta,t_2)
$$
To avoid the secular increase of $v_1$ we obtain that $\pt_{\zeta\eta}^2v_1 = 0$. It allows us to derive equations for the leading-order terms
\begin{eqnarray*}
\pt_{t_2\zeta}^2v_0^+ + a \pt_{\zeta}v_0^+\pt_{\zeta}^2v_0^+ &=& 0,  \\
\pt_{t_2\eta}^2v_0^- + a \pt_{\eta}v_0^-\pt_{\eta}^2v_0^- &=&0. 
\end{eqnarray*}
By substitution $V^+=\pt_{\zeta}v_0^+$ and $V^-=\pt_{\eta}v_0^-$ we get the pair of the Hopf equations
\begin{eqnarray}
\pt_{t_2}V^+ + a V^+\pt_{\zeta}V^+ &=& 0, \nonumber \\
\pt_{t_2}V^- + a V^- \pt_{\eta}V^- &=&0. \label{Hopf}
\end{eqnarray} 
Functions $v_1^+, v_1^-, v_2^+, v_2^-, v_2$ are determined in much the same way.

\subsection{The domain of validity for (\ref{secondExternalExpansion})}

The usual condition for validity
$$
\ve \max_{x_1,t_1,x_2,t_2}\vert v_{n+1}\vert = o \left(\max_{x_1,t_1,x_2,t_2}\vert v_{n}\vert\right), \quad \ve \to 0
$$
yields 
$$
t_2 \gg \ve.
$$
\par
\section*{Concluding remarks}

The realized rough analytical and numerical calculations allow one to obtain the finite amplitude waves due to the slow passage through the resonance of the small driving force. 
As it was mentioned above this large increase of the amplitude takes place due to the resonance on the zero harmonic. It becomes clear that  this  approach is evaluable for the phase $S(x,t)$ of  more general type.  The construction and matching of infinite asymptotic series are  possible also. The such type representation of the solution for the original problem opens the way for a justification of the constructed asymptotics. But this is a theme for our future investigations.
\par

{\bf Acknowledgements\,}.  This article was partially written during the stay of the third  author at the Institute of Mathematics, University of Potsdam. He gratefully acknowledges the financial support of the Deuthcher Akademischer Austauschdienst. The work was also supported by grants RFBR 03-01-00716, Leading Scientific Schools 1446.2003.1 and INTAS 03-51-4286.
\par

\end{document}